\documentclass[twocolumn,showpacs,preprintnumbers,amsmath,amssymb,prl]{revtex4-1}

\usepackage{color}
\usepackage{graphicx}
\usepackage{dcolumn}
\usepackage{bm}
\usepackage{soul}

\begin{document}


\title{Sideband Rabi spectroscopy of finite-temperature trapped Bose gases}
\author{Baptiste Allard}
\author{Matteo Fadel}
\author{Roman Schmied}
\author{Philipp Treutlein}\email{philipp.treutlein@unibas.ch}

\affiliation{Department of Physics, University of Basel, Klingelbergstrasse 82, 4056 Basel, Switzerland}

\date{\today}

\begin{abstract}
We use Rabi spectroscopy to explore the low-energy excitation spectrum of a finite-temperature Bose gas of rubidium atoms across the phase transition to a Bose-Einstein condensate (BEC). To record this spectrum, we coherently drive the atomic population between two spin states. A small relative displacement of the spin-specific trapping potentials enables sideband transitions between different motional states.
The intrinsic non-linearity of the motional spectrum, mainly originating from two-body interactions, 
makes it possible to resolve and address individual excitation lines. Together with sensitive atom-counting, this constitutes a feasible technique to count single excited atoms of a BEC and to determine the temperature of nearly pure condensates. As an example, we show that for a nearly pure BEC of $N=800$ atoms the first excited state has a population of less than $5$ atoms, corresponding to an upper bound on the temperature of $30~\mathrm{nK}$. 
\end{abstract}

\pacs{03.75.Kk, 67.85.De, 82.53.Kp, 37.10.Vz}

\maketitle

Experimental realizations of atomic Bose-Einstein condensates (BECs) are always at finite (\textit{i.e.}\ non-zero) temperature, meaning that excited states are also populated. Besides being interesting from a fundamental point of view, this is also expected to limit the coherence time of the BECs \cite{Sinatra09, Sinatra13a} and the fidelity of non-classical states \cite{Sinatra11, Sinatra13b}, posing a limit to their application in quantum metrology and quantum information processing. Experimentally, this has been explored in particular in the case of BECs in a double-well potential \cite{Gati06, Hofferberth08}. 

For the study of finite-temperature effects, a key ingredient is the knowledge of the excitation spectrum of the system, from which thermodynamic quantities can be derived. For this reason, the excitation spectrum of a BEC has been a matter of extensive studies \cite{Dalfovo99}. Experimentally one relies both on the ability of preparing BECs with the desired particle number and temperature, and on addressing specific excitation lines. For this, mainly two techniques have been developed and applied to various temperature regimes: trapping-potential perturbations to study low-energy excitations \citep{Jin96,Mewes96,Jin97}, and two-photon Bragg transitions to probe high-energy excitations (where the spectrum is a quasi-continuum) \cite{Stenger99,Stamper99,Ozeri05}. 

In contrast, experiments with single particles or non-interacting ensembles in ion traps and optical dipole traps have successfully adopted sideband-resolved Raman and Rabi spectroscopy techniques to probe the particle's excitation spectrum and control its external degrees of freedom \cite{Diedrich89,Wineland98,Boozer06,Kaufman12,Foerster09}. In particular, these techniques can be used to measure the temperature or to perform motional ground-state cooling in systems where the lowest part of the spectrum has modes equally spaced in energy \cite{Boozer06,Kaufman12,Foerster09}.

\begin{figure}
\includegraphics[width=0.5\columnwidth]{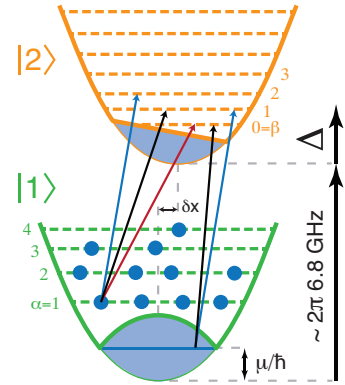}
\caption{\label{scheme}(Color online) Rabi spectroscopy of trapped Bose gases. Green (orange) thick solid lines sketch the effective trapping potential for single-particle excitations in state $\vert 1\rangle$ ($\vert 2\rangle$). The blue shaded regions represent mean-field potentials. Horizontal dashed lines represent the single-particle energy levels. The few first values $\alpha$ and $\beta$ of the indices used in Eqs.~\eqref{ZT2} and \eqref{ZT3} are shown for each trap. The blue horizontal line indicates the BEC chemical potential (see Eq.~\eqref{ZT1}). The arrows show some possible transitions from the BEC or from the thermal cloud (blue dots). Black arrows are called carrier transitions while blue (red) arrows are examples of blue (red) sideband transitions.}
\end{figure}

In this paper, we extend the Rabi spectroscopy technique to an interacting many-body system, to probe the excitation spectrum of a finite-temperature Bose gas in a harmonic trap across the BEC phase transition. Moreover, we present a method to control the temperature at fixed atom number in such a system. The experimental procedure is similar to the one adopted in Ref.~\onlinecite{Foerster09} for non-interacting thermal atoms. We apply a Rabi pulse that coherently couples two atomic internal states, whose trapping potentials are slightly spatially displaced by a microwave dressed-state potential \cite{Boehi09} to enable transitions between different motional states. Then, final populations of the two internal states are measured as functions of the frequency of the Rabi drive to characterize the excitation spectrum of the finite-temperature sample across the BEC phase transition. Above the critical temperature, we observe almost symmetric sidebands, while at extremely low temperatures no red sideband can be detected, placing a lower bound on the number of particles in excited modes in a region where standard time-of-flight measurements are ineffective. 

\begin{figure}[!ht]
\includegraphics[width=\columnwidth]{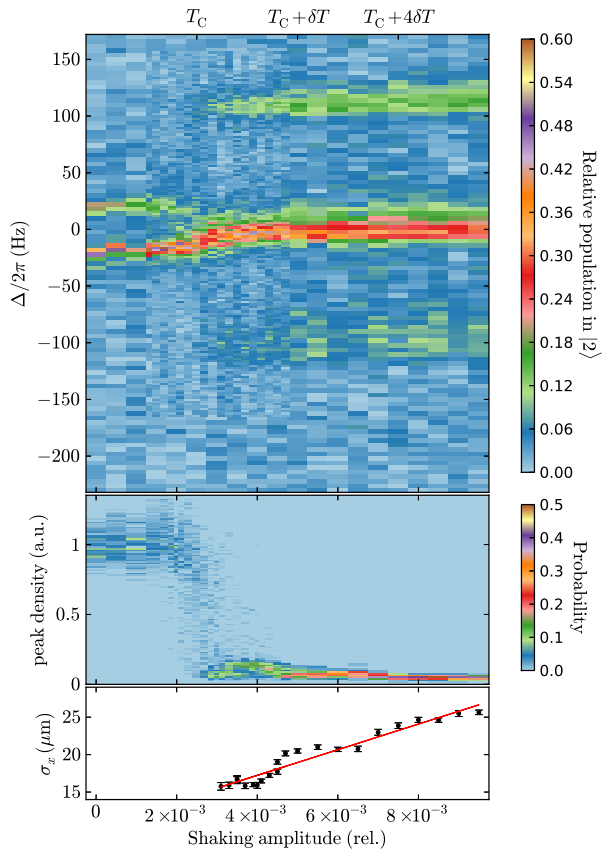}
\caption{\label{spect-vs-T_0p13}(Color online) Finite temperature spectroscopy for $\delta x = 0.13~\mu\mathrm{m}$ and $N=400$ atoms. Top panel: Population transfer as a function of the shaking amplitude and the detuning $\Delta$. The $\Delta=0$ detuning point is identified with the center of the carrier line for the non-interacting thermal ensemble. Central panel: histograms of the peak density after time-of-flight as a function of the shaking amplitude. The peak density is roughly proportional to the condensed fraction. It is rescaled such that $0$ corresponds to no atoms and $1$ is reached on average for the pure BEC. $T_\mathrm{C}$ is identified with the location of the step, and corresponds to a relative shaking of $2.5\times 10^{-3}$. Bottom panel : Fitted Gaussian widths of the thermal distribution as a function of the shaking amplitude. Black points are statistical averages; error bars are errors on the mean. No data are shown for small shaking amplitudes since the clouds do not show any thermal component. The red line is a linear fit to the data, showing a quadratic scaling of the temperature measured in time-of-flight with the shaking amplitude for $T>T_{\mathrm{C}}$. Since our time-of-flight sequence starts with a smooth release ramp to optimize atom counting, we cannot extract any quantitative in-trap temperature estimates from the fitted widths, but the observed scaling with the shaking amplitude holds nonetheless. The in-trap temperature axis drawn on top of the figure is constructed from the known scaling with the shaking amplitude for a thermal ensemble and from the location of the critical point.}
\end{figure}

We use $^{87}$Rb atoms in the hyperfine states $\vert 1 \rangle \equiv \vert F=1, m_{\mathrm{F}}=-1\rangle$ and $\vert 2 \rangle \equiv \vert F=2, m_{\mathrm{F}}=1\rangle$, confined in a chip-based magnetic trap characterized by its oscillation frequencies $\omega_x/2\pi=112~\mathrm{Hz}$ along the axial direction and $\omega_y/2\pi=\omega_z/2\pi=517~\mathrm{Hz}$ along the transverse directions. Because the magnetic moments of $\vert 1\rangle$ and $\vert 2 \rangle$ are approximately the same for magnetic fields near $3.23~\mathrm{G}$, the same trap is experienced by both states \cite{Treutlein04, Harber02}. To allow for independent control of both atom number and temperature, we use the following sequence. We first prepare a pure BEC containing several thousand particles in $\vert 1 \rangle$ by radio-frequency (rf) evaporative cooling \cite{Ketterle99}. Atom number and temperature of such a BEC are not independent. To overcome this, a further rf evaporation sets and stabilizes the temperature and the particle number (to $N=400 \pm 50$) by cutting deeply into the BEC with the ``rf-knife''. Then, the temperature of the sample is increased by applying a controlled sinusoidal shaking of the trap. The current in the wire that provides transverse confinement of the atoms \cite{Boehi09} is slightly modulated at $\omega_{\mathrm{mod}}$ during $400~\mathrm{ms}$. Primarily, this results in an oscillation of the trap position along the vertical axis $z$ with an amplitude of $\pm 4~\mu\mathrm{m}$ for a current modulation with relative amplitude of $1.0\times 10^{-2}$. The amplitude along the two other directions is of the order of $1~\mu\mathrm{m}$, and the axial (transverse) trap frequency is modulated by up to $16\%$ ($8\%$). The modulation frequency is set to $\omega_{\mathrm{mod}}=2\pi \times 1035~\mathrm{Hz}\simeq 2\omega_z\simeq 2\omega_y$, and heats the sample by resonant parametric excitation of the transverse motion. After the shaking, we wait for $1~\mathrm{s}$ to ensure thermalization of the cloud. This has been verified by noting that the oscillation of both shape and position of the atom cloud in time-of-flight measurements are damped, and it is consistent with estimates of the thermalization time due to collisions at our densities \cite{Dalfovo99}. For a parametric heating process \cite{Gehm98}, we expect that the deposited energy scales with the squared modulation amplitude. While the heat capacity of a thermal cloud is constant, that of a BEC is non-trivial \cite{Gati06}. In Fig.~\ref{spect-vs-T_0p13}, this leads to a distorted temperature scale below $T_{\mathrm{C}}$. By time-of-flight images, we observe that for samples in which thermal wings are visible, the width of these wings increases linearly with the shaking amplitude, confirming that the temperature in time-of-flight scales as the square of the shaking amplitude (see bottom panel of fig.~\ref{spect-vs-T_0p13}). 

Our spectroscopy pulse is a two-photon Rabi drive, coupling $\vert 1 \rangle$ to $\vert 2 \rangle$ (see Fig.~\ref{scheme} and Refs.~\onlinecite{Treutlein04,Boehi09}). The driving fields are emitted by an rf-coil and a microwave horn, placed far away from the chip so that the fields are spatially homogeneous over the sample. While the frequency of the microwave is kept fixed, the frequency of the radio wave can be changed to vary the detuning $\Delta$ of the Rabi pulse with respect to the $\vert 1 \rangle$ to $\vert 2 \rangle$ transition. $\Delta=0$ is defined as the resonance of the Rabi drive with the trap bottoms, taking microwave level shifts into account (see Fig.~\ref{scheme}). The Rabi pulse is weak in the sense that the Rabi frequency $\Omega_{\mathrm{Rabi}}/2\pi \simeq 3.5~\mathrm{Hz}$ is much smaller than the trapping frequencies. The Rabi pulse length is $t_{\mathrm{Rabi}}=140~\mathrm{ms}$ corresponding to a full spectral width at half maximum of the order of $10~\mathrm{Hz}$, sufficient to resolve the vibrational sidebands in the trap. 

If the traps for the two states $\vert 1\rangle$ and $\vert 2\rangle$ are perfectly identical (\textit{i.e.}\ $\delta x = 0$ in Fig.~\ref{scheme}), the only possible vibrational transitions are carrier transitions (black arrows in Fig.~\ref{scheme}). To enable sideband transitions (blue and red arrows in Fig.~\ref{scheme}), we apply a state-dependent microwave potential \cite{Boehi09} generated by an on-chip microwave current. It displaces the trap for $\vert 2\rangle$ by $\delta x$, much less than the typical size of the BEC, along its weak confinement axis $x$ (see Fig.~\ref{scheme} and Ref.~\onlinecite{Boehi09}). This state-dependent potential also decreases the energy difference between the two trap bottoms by a few hundred Hz due to microwave level shifts, which we take into account in our definition of $\Delta$.

After the spectroscopy pulse, the final populations are recorded with two absorption images, which allow us to independently count the number of particles transferred by the pulse into $\vert 2\rangle$ and the remaining particles in $\vert1\rangle$, in every single realization of the experiment. Before detection we release the atoms by smoothly ramping down the trapping potential so that the atom cloud is decompressed and accelerated away from the chip surface. This release ramp is optimized so that the cloud is small at the time of detection, maximizing the atom number sensitivity. This decompression produces an adiabatic cooling that reduces the observed temperature in time-of-flight compared to the in-trap temperature by a constant factor. The scaling we observe with the shaking amplitude is thus valid for in-trap temperatures as well.

The top panel of Fig.~\ref{spect-vs-T_0p13} shows transfer probabilities (relative population in state $\vert 2\rangle$ after the Rabi pulse) as a function of the detuning $\Delta$ and the trap shaking amplitude, for a displacement between the two traps of $\delta x = 0.13~\mu\mathrm{m}$. Note that the maximal population transfer stays always below $50\%$, even if $\Omega_{\mathrm{Rabi}}t_{\mathrm{Rabi}}\simeq \pi$. This is because even if the pulse starts on resonance with some transition, it will end up off resonance during the population transfer due to changes in the mean-field potential \footnote{Because $\Omega_{\mathrm{Rabi}}\ll \omega_{x,y,z}$, the excited-state energies and mode functions change adiabatically as the mean-field potential changes and do not acquire a spatial dependence.}. This panel clearly shows two distinct behaviors. On the left side, \textit{i.e.}\ on the low temperature/pure BEC side, one carrier line and one blue sideband are visible in the spectrum. The carrier line (blue sideband) corresponds to population transfer from the BEC in $\vert 1\rangle$ to the ground state (first excited state) in the effective trapping potential for $\vert 2\rangle$ (see Fig.~\ref{scheme}). There is no visible red sideband signal, confirming that the BEC is very pure in our coldest sample. For a non-interacting system, we would expect that the first blue sideband is separated from the carrier by $\omega_x/2\pi=112~\mathrm{Hz}$. Here, we measure $42~\mathrm{Hz}$, which means that the inter- and intra-species interactions strongly affect the spectrum on the BEC side.
On the right side of Fig.~\ref{spect-vs-T_0p13}, \textit{i.e.}\ on the high temperature side \footnote{Time of flight images confirm that the temperature of such a heated cloud is above the BEC transition temperature.}, the spectrum shows both red and blue sidebands surronding the carrier line at $\approx\pm \omega_x / 2\pi$, close to what one would expect for a non-interacting ensemble.

Between these two extreme cases, the system crosses the BEC critical point. From the sudden decrease of the peak density (central panel of Fig.~\ref{spect-vs-T_0p13}), as well as from the dramatic change observed in the spectrum, we estimate that the sample reaches the critical temperature $T_\mathrm{C}$ for a relative shaking amplitude of $2.5\times 10^{-3}$. For our atom number ($N=400$) and trapping potential, considering finite size and interaction effects \cite{Dalfovo99}, we calculate $T_{\mathrm{C}}=87~\mathrm{nK}$. Around the critical point, the peak density fluctuates strongly between a low value, characteristic of a thermal ensemble, and a high value, indicating the existence of a BEC. Approching $T_{\mathrm{C}}$ from higher temperatures, we can see that the frequencies of the sidebands stay constant while the amplitude gradually decreases and the amplitude asymmetry becomes more pronounced. This asymmetry is stronger than predicted by the Bose-Einstein distribution in a harmonic trap, and it is explained below by a model that includes trap anharmonicities. At temperatures below $T_{\mathrm{C}}$ the thermal sidebands are no longer visible, and the two-peak structure appears, indicating that transitions from the macroscopically-populated BEC mode dominate the spectrum. As the condensate fraction increases, the carrier line bends down due to mean-field shifts. The total shift of the carrier line (from the non-interacting/high temperature limit to the fully condensed one) is $\delta\!f\simeq22~\mathrm{Hz}$. To first approximation, if we neglect the displacement between the two traps and assume a small transfer, this shift can be seen as the energy difference between a BEC in $\vert 1\rangle$ containing $N$ particles and a BEC of $N-1$ particles in $\vert 1\rangle$ interacting with a single particle in the same spatial mode but in $\vert 2 \rangle$. This energy difference is $h\: \delta\!f = \mu\left(a_{12}/a_{11}-1\right)$, where $a_{11}=100.4\,a_0$ and $a_{12}=98.01\,a_0$ are the scattering lengths \cite{Egorov13}, $a_0$ is the Bohr radius, and $\mu$ is the chemical potential in $\vert 1 \rangle$. Solving for $\delta\!f = 22~\mathrm{Hz}$, we expect the chemical potential to be $\mu/h=922~\mathrm{Hz}$. A Gross-Pitaevskii simulation of our system gives $\mu/h=966~\mathrm{Hz}$, in agreement with the measurement.

\begin{figure}[!ht]
\includegraphics[width=\columnwidth]{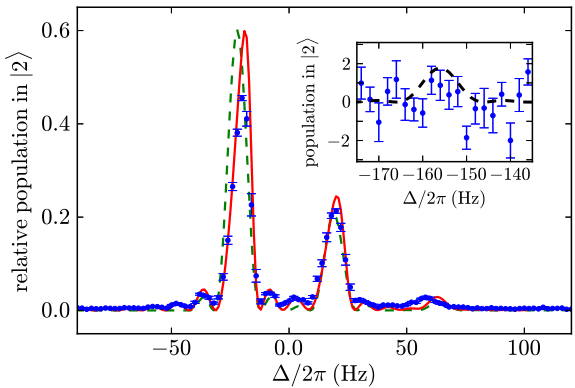}
\caption{\label{bec_sim-exp}(Color online) Rabi sideband spectroscopy (blue points) for a pure BEC of $N=800$ atoms as a function of the detuning, corresponding to the leftmost column of the top panel of Fig.~\ref{spect-vs-T_0p13}. The errorbars indicate statistical uncertainties. The red continuous line is calculated with a time-dependent two-component Gross-Pitaevskii model \eqref{tdepGPE}. The green dashed line is the result of the time-independent model described in the text, Eqs.~\eqref{ZT1}-\eqref{ZT3}, at zero-temperature. Inset: Zoom on the red sideband region plotted in units of detected atoms in $\vert 2\rangle$. The number of excitations transferred is consistent with zero, within our detection uncertainty. The dashed black line is the simulated spectrum (see text) expected for a BEC containing $5$ atoms in the first excited state, corresponding in our case to a temperature of $30~\mathrm{nK}$, or $T/T_{\mathrm{C}}=0.27$ for $N=800$.}
\end{figure}

A detailed spectrum of the pure BEC case is shown in Fig.~\ref{bec_sim-exp}, corresponding to the leftmost column of the top panel of Fig.~\ref{spect-vs-T_0p13} but in this case recorded with $N=800$. The measurement is compared to a time-dependent two-component Gross-Pitaevskii model (GPM) \cite{Cirac98} in the rotating-wave approximation (red line) taking as inputs the atom number, the trapping potentials ($V_1$ and $V_2$, assumed harmonic), the splitting distance $\delta x$ and the Rabi pulse length $t_{\mathrm{Rabi}}$ and strength $\Omega_\mathrm{Rabi}$ calibrated from independent measurements. The coupled equations for the two BEC wavefunctions $\Psi_1$ and $\Psi_2$ are

\begin{eqnarray}\label{tdepGPE}
\nonumber i\hbar \frac{\partial }{\partial t} \Psi_1\ =&&-\frac{\hbar^{2}}{2 m} \nabla^{2} \Psi_1 + \left[V_1 + g_{11}\vert\Psi_1\vert^{2}+g_{12}\vert\Psi_2\vert^{2} \right]\Psi_1 \\
&&+ \frac{\hbar\Omega_\mathrm{Rabi}}{2} e^{-i\Delta t}\Psi_2 , \nonumber\\
\nonumber i\hbar \frac{\partial }{\partial t} \Psi_2\ =&&-\frac{\hbar^{2}}{2 m} \nabla^{2} \Psi_2 + \left[V_2 + g_{22}\vert\Psi_2\vert^{2}+g_{12}\vert\Psi_1\vert^{2} \right]\Psi_2 \\ 
&&+ \frac{\hbar\Omega_\mathrm{Rabi}}{2} e^{i\Delta t}\Psi_1 ,
\end{eqnarray}
where $m$ is the atomic mass, $\Delta$ is the detuning of the Rabi pulse with respect to the $\vert 1\rangle$  to $\vert 2 \rangle$ transition, $g_{ij}=4\pi\hbar^{2}a_{ij}/m$ are the collisional interaction strengths (with $a_{22}=95.44 a_0$), and the wavefunctions are normalized to the particle numbers in each mode. The two trapping potentials are identical but separated by $\delta x$, \textit{i.e.}\ $V_2(\vec{r})=V_1(\vec{r}-\delta x \,\vec{e}_x)$. A simulation of the trapping potential shows that the trap anharmonicity can be neglected for the pure BEC case. Our GPM neglects particle losses \cite{Egorov13}, since we have noticed that the spectrum is insensitive to losses for the evolution time we are considering. Without any free parameters, the GPM reproduces the width of the lines, the spacing between them and their strength. Note that the slight overestimation of the strength of the carrier line can be explained by a miscalibration of both the pulse area and the displacement $\delta x$.

With infinitely slow Rabi drive and infinitesimal transfer, the spectrum at zero temperature should present several sharp lines, one per eigenstate for $\vert 2\rangle$ weighted by the Franck-Condon factor with the initial BEC wavefunction. In practice, however, the peaks are convolved with the Fourier transform of the finite-time Rabi pulse and distorted by the Rabi frequency nonlinearity. The distorted cardinal sine (sinc) shape observed in Fig.~\ref{bec_sim-exp} for both the experiment and the model confirms that the Rabi pulse is Fourier limited and that the time-dependent GPM is needed for a fine analysis of the spectrum.

We also compare the spectrum with the following time-independent model \cite{Oehberg97}, which allows to describe also finite-temperature cases. For state $\vert 1\rangle$, the BEC mode function $\Psi_1$ is obtained from the Gross-Pitaevskii equation,

\begin{equation}
\label{ZT1}
\mu\Psi_1\ =\left[-\frac{\hbar^{2}}{2 m} \nabla^{2}  + V_1 + g_{11}\vert\Psi_1\vert^{2}\right]\Psi_1. 
\end{equation}
In the Hartree-Fock approximation, single-particle excitations $\Psi_{1\alpha}$ are obtained by diagonalizing 
\begin{equation}
\label{ZT2}
E_{1\alpha}\Psi_{1\alpha}\ =\left[-\frac{\hbar^{2}}{2 m} \nabla^{2}  + V_1 +2 g_{11}\vert\Psi_1\vert^{2}\right]\Psi_{1\alpha}.
\end{equation}
For state $\vert 2 \rangle$, the single-particle excitations $\Psi_{2\beta}$ satisfy
\begin{equation}
\label{ZT3}
E_{2\beta}\Psi_{2\beta}\ =\left[-\frac{\hbar^{2}}{2 m} \nabla^{2}  + V_2 + g_{12}\vert\Psi_1\vert^{2}\right]\Psi_{2\beta}. 
\end{equation}
Note that $\alpha$ and $\beta$ are non-negative integers, labeling  all vibrational states of the full 3D model (see fig.~\ref{scheme}).

Since at zero temperature the only initially populated state is the BEC in $ \Psi_1$, all the allowed transitions have energy differences $\Delta E_{\beta} = E_{2\beta}-\mu$ and are weighted by $\vert\langle\Psi_1\vert\Psi_{2\beta}\rangle \vert^{2}$, according to Fermi's golden rule. To reproduce the finite spectral size of the Rabi pulse, we convolve the spectrum with a sinc function of $10~\mathrm{Hz}$ FWHM. Since this model does not predict the absolute transfer efficiency, we set the overall amplitude of the spectrum to be the same as for the time-dependent GPM. 
The result is shown in Fig.~\ref{bec_sim-exp} as the green-dashed line. The model reproduces the data well, with the exception of the asymmetric shape of the peaks.

\begin{figure}[!ht]
\includegraphics[width=\columnwidth]{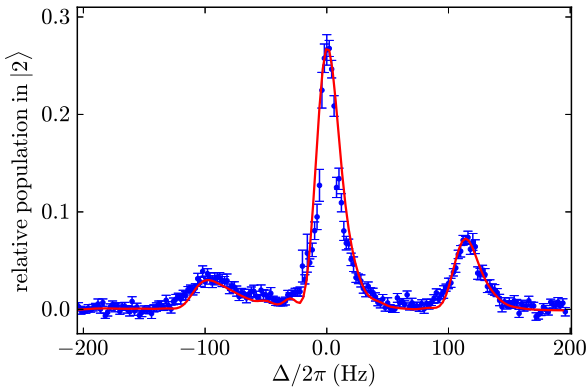}
\caption{\label{th_HF}(Color online) Rabi sideband spectroscopy (blue points) for a thermal ensemble as a function of the detuning ($5.0\times 10^{-3}$ relative shaking amplitude). The errorbars indicate statistical uncertainties. The red line is the result from the high-temperature model described in the text, see Eqs.~\eqref{HTmodel}. We find $\gamma \simeq 2\pi \times 2.5~\mathrm{Hz~\mu m^{-1}}$ .}
\end{figure}

At non-zero temperatures, this time-independent model predicts also transitions from the thermal population with energy $\Delta E_{\alpha,\beta} = E_{2\beta}-E_{1\alpha}$ and weighted by $\vert\langle\Psi_{1\alpha}\vert\Psi_{2\beta}\rangle \vert^{2}$. Among all these transitions, the strongest one is expected for $\alpha=\beta=1$ and should be seen around $\Delta/2\pi=-160~\mathrm{Hz}$, about $140~\mathrm{Hz}$ below the carrier line \footnote{There is another line corresponding to the transition from the first excited vibrational state in \unexpanded{$\vert 1 \rangle$} ($\alpha=1$) to the lowest vibrational state in \unexpanded{$\vert 2 \rangle$}($\beta=0$) at an offset of $-240~\text{Hz}$ from the carrier. This transition is weaker because it has a lower wavefunction overlap. We focus our analysis on the strongest transition involving the thermal cloud.}.  The inset in Fig.~\ref{bec_sim-exp} shows a zoom into this region of the spectrum, but in units of atoms detected in state $\vert 2\rangle$. Our imaging system has a detection noise of $\approx 4$ atoms for a single-shot measurement. The error bars in Fig.~\ref{bec_sim-exp} take into account that we averaged over $20$ measurements per detuning. The measured population in $\vert 2\rangle$ at the expected frequency of the red sideband is consistent with zero within our measurement uncertainty. We estimate the transfer probability at this red sideband to $20\%$. We can thus put an upper bound on the population of the first excited vibrational state in our coldest sample of $\approx 5$. This estimate is confirmed by the simulation of the spectrum of a BEC containing $5$ atoms in the first excited state, corresponding to a temperature of $30~\mathrm{nK}$, (dashed black line in the inset of Fig.~\ref{bec_sim-exp}). For these parameters, the total number of thermal excitations is $15$. This spectrum is computed from the time-independent model Eqs.~(\ref{ZT1})-(\ref{ZT3}), where the populations of the excited states are given by the Bose-Einstein distribution. Detection of single excitations could be achieved in our setup by averaging over $1500$ measurements.

A detailed spectrum for a thermal ensemble with $N\approx 800$ and for a reative shaking amplitude of $5.0\times 10^{-3}$ is shown in Fig.~\ref{th_HF}. In the case of non-interacting particles in a harmonic trap, the spectrum would show three lines spaced by $\omega_x/2\pi$ convolved with the Fourier transform of the Rabi pulse. The main difference to this ideal picture is that the peaks we observe are asymmetric, not equally spaced and broader than expected. This can be explained by an anharmonicity in the spectrum, which can arise from interactions and anharmonicities in the trapping potentials. To describe this high temperature system, we use the following model. Thermal atoms in $\vert 1\rangle$ ($\vert 2\rangle$) occupy the modes $\Psi_{1\alpha}$ ($\Psi_{2\beta}$), solutions of the Schr\"odinger equations 

\begin{eqnarray}\label{HTmodel}
E_{1\alpha}\Psi_{1\alpha}\ =&&\left[-\frac{\hbar^{2}}{2 m} \nabla^{2}  + V_1 + \frac{1}{2} m \gamma^2 x^4 \right]\Psi_{1\alpha},\\
E_{2\beta}\Psi_{2\beta}\ =&&\left[-\frac{\hbar^{2}}{2 m} \nabla^{2}  + V_2 + \frac{1}{2} m  \gamma^2 \left(x-\delta x\right)^4 \right]\Psi_{2\beta},\nonumber
\end{eqnarray}
where the constant $\gamma$ characterizes the potential anharmonicity. In practice, it is a free parameter extracted by fitting the data. We found $\gamma \simeq 2\pi \times 2.5~\mathrm{Hz~\mu m}^{-1}$, which we check to be in reasonable agreement with expected anharmonicity from simulations. Like in the previous model, the spectrum we obtain is the sum over all allowed transitions between a thermally populated $\Psi_{1\alpha}$ and an initially empty mode $\Psi_{2\beta}$. 
We can see from Fig.~\ref{th_HF} that this model (red line in figure) gives a very good description of the regime above $T_{\mathrm{C}}$. It is interesting to note that the relative amplitude between the red and the blue sideband is strongly affected by anharmonicities, making it difficult and unreliable to extract a temperature from it.

\begin{figure}[!ht]
\includegraphics[width=\columnwidth]{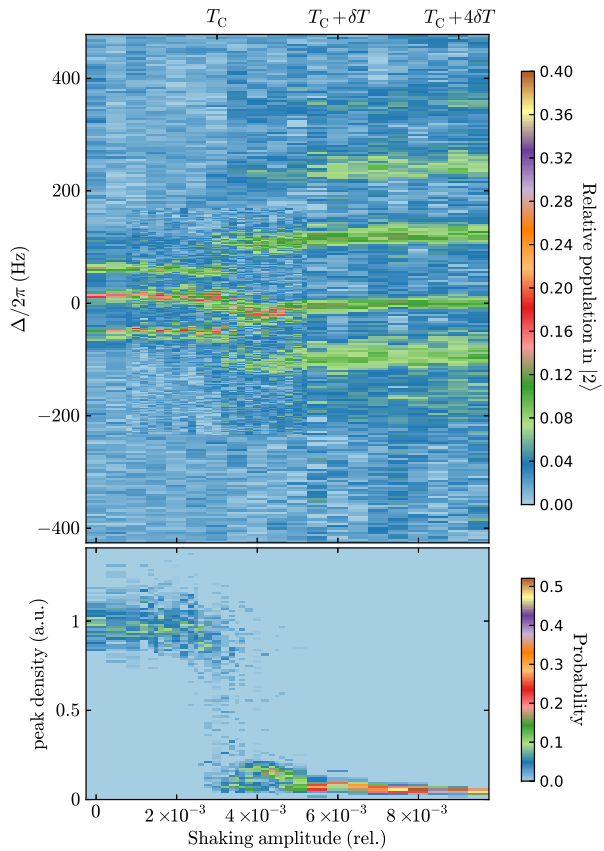}
\caption{\label{spect-vs-T_0p26}(Color online) Finite temperature spectroscopy for $\delta x = 0.26~\mu\mathrm{m}$. Same experiment as in Fig.~\ref{spect-vs-T_0p13} with twice the splitting distance. Here the BEC transition temperature is identified at a relative shaking amplitude of $3.0\times 10^{-3}$ consistent with a slightly higher atom number $N=630$ for this dataset.}
\end{figure}

For comparison, Fig.~\ref{spect-vs-T_0p26} shows the same experiment as in Fig.~\ref{spect-vs-T_0p13} but for a splitting distance $\delta x=0.26~\mu\mathrm{m}$ and $N=630$. A larger splitting distance increases the overlap between eigenstates with larger difference of vibrational index, allowing higher-order Rabi transitions. On the BEC side a second blue sideband appears and, in the low temperature limit, its position is also in good agreement with the models Eq.~(\ref{tdepGPE}) and Eqs.~(\ref{ZT1})-(\ref{ZT3}). On the high temperature side, three blue sidebands and at least two red sidebands are observed. An additional feature visible on this side is the gradual appearance of higher-order sidebands as the temperature increases. It is interesting to note that, for the temperature range we are probing, the ratio between the amplitudes of a blue sideband and its red counterpart has a weaker temperature dependence than the ratio between the amplitudes of a blue (or red) sideband and an adjacent one. However, already the simple model of a single particle in a harmonic potential shows that the former ratio is better suited for thermometry, since the only free parameters are the trapping frequencies, while the latter ratio also depends on the Franck-Condon factor, and therefore the splitting distance is needed in order to extract the temperature.

In conclusion, we presented a technique to prepare BECs with a well-defined atom number and variable temperature, using a rf-evaporation and a subsequent controlled shaking of the trap, and the sideband Rabi spectroscopy of such finite-temperature BECs. The measured spectra show resonance peaks that are in good agreement with theoretical models. This spectroscopy technique is especially efficient to probe the low energy excitations. In the coldest BECs we prepared, we can exclude the presence of more than $5$ atoms in the first excited state, corresponding to a temperature below $30~\mathrm{nK}$. At this temperature, no thermal cloud is visible in time-of-flight images. We estimate that single excitation detection is reachable in our setup with $1500$ measurements per detuning \footnote{We point out that in the specific case of a BEC in a double-well potential, a primary noise thermometer is presented in Ref~\onlinecite{Gati06}.}. A higher sensitivity per experimental realization can be reached with a single-atom resolution imaging system, and by optimizing the Rabi pulse parameters. Furthermore, this technique can be used to prepare BECs in excited motional states.

We acknowledge helpful discussions with A.~Sinatra, C.~Klempt, and H.-C.~N\"agerl. This work was supported by the Swiss National Science Foundation.

\bibliographystyle{apsrev4-1}
\bibliography{spectro}

\begin{thebibliography}{30}%
\makeatletter
\providecommand \@ifxundefined [1]{%
 \@ifx{#1\undefined}
}%
\providecommand \@ifnum [1]{%
 \ifnum #1\expandafter \@firstoftwo
 \else \expandafter \@secondoftwo
 \fi
}%
\providecommand \@ifx [1]{%
 \ifx #1\expandafter \@firstoftwo
 \else \expandafter \@secondoftwo
 \fi
}%
\providecommand \natexlab [1]{#1}%
\providecommand \enquote  [1]{``#1''}%
\providecommand \bibnamefont  [1]{#1}%
\providecommand \bibfnamefont [1]{#1}%
\providecommand \citenamefont [1]{#1}%
\providecommand \href@noop [0]{\@secondoftwo}%
\providecommand \href [0]{\begingroup \@sanitize@url \@href}%
\providecommand \@href[1]{\@@startlink{#1}\@@href}%
\providecommand \@@href[1]{\endgroup#1\@@endlink}%
\providecommand \@sanitize@url [0]{\catcode `\\12\catcode `\$12\catcode
  `\&12\catcode `\#12\catcode `\^12\catcode `\_12\catcode `\%12\relax}%
\providecommand \@@startlink[1]{}%
\providecommand \@@endlink[0]{}%
\providecommand \url  [0]{\begingroup\@sanitize@url \@url }%
\providecommand \@url [1]{\endgroup\@href {#1}{\urlprefix }}%
\providecommand \urlprefix  [0]{URL }%
\providecommand \Eprint [0]{\href }%
\providecommand \doibase [0]{http://dx.doi.org/}%
\providecommand \selectlanguage [0]{\@gobble}%
\providecommand \bibinfo  [0]{\@secondoftwo}%
\providecommand \bibfield  [0]{\@secondoftwo}%
\providecommand \translation [1]{[#1]}%
\providecommand \BibitemOpen [0]{}%
\providecommand \bibitemStop [0]{}%
\providecommand \bibitemNoStop [0]{.\EOS\space}%
\providecommand \EOS [0]{\spacefactor3000\relax}%
\providecommand \BibitemShut  [1]{\csname bibitem#1\endcsname}%
\let\auto@bib@innerbib\@empty
\bibitem [{\citenamefont {Sinatra}\ \emph {et~al.}(2009)\citenamefont
  {Sinatra}, \citenamefont {Castin},\ and\ \citenamefont
  {Witkowska}}]{Sinatra09}%
  \BibitemOpen
  \bibfield  {author} {\bibinfo {author} {\bibfnamefont {A.}~\bibnamefont
  {Sinatra}}, \bibinfo {author} {\bibfnamefont {Y.}~\bibnamefont {Castin}}, \
  and\ \bibinfo {author} {\bibfnamefont {E.}~\bibnamefont {Witkowska}},\ }\href
  {\doibase 10.1103/PhysRevA.80.033614} {\bibfield  {journal} {\bibinfo
  {journal} {Phys. Rev. A}\ }\textbf {\bibinfo {volume} {80}},\ \bibinfo
  {pages} {033614} (\bibinfo {year} {2009})}\BibitemShut {NoStop}%
\bibitem [{\citenamefont {Sinatra}\ and\ \citenamefont
  {Castin}(2013)}]{Sinatra13a}%
  \BibitemOpen
  \bibfield  {author} {\bibinfo {author} {\bibfnamefont {A.}~\bibnamefont
  {Sinatra}}\ and\ \bibinfo {author} {\bibfnamefont {Y.}~\bibnamefont
  {Castin}},\ }\enquote {\bibinfo {title} {Physics of quantum fluids: {N}ew
  trends and hot topics in atomic and polariton condensates},}\ \ (\bibinfo
  {publisher} {Springer},\ \bibinfo {year} {2013})\ Chap.\ \bibinfo {chapter}
  {Spatial and temporal coherence of a {B}ose-condensed gas}, pp.\ \bibinfo
  {pages} {315--339}\BibitemShut {NoStop}%
\bibitem [{\citenamefont {Sinatra}\ \emph {et~al.}(2011)\citenamefont
  {Sinatra}, \citenamefont {Witkowska}, \citenamefont {Dornstetter},
  \citenamefont {Li},\ and\ \citenamefont {Castin}}]{Sinatra11}%
  \BibitemOpen
  \bibfield  {author} {\bibinfo {author} {\bibfnamefont {A.}~\bibnamefont
  {Sinatra}}, \bibinfo {author} {\bibfnamefont {E.}~\bibnamefont {Witkowska}},
  \bibinfo {author} {\bibfnamefont {J.-C.}\ \bibnamefont {Dornstetter}},
  \bibinfo {author} {\bibfnamefont {Y.}~\bibnamefont {Li}}, \ and\ \bibinfo
  {author} {\bibfnamefont {Y.}~\bibnamefont {Castin}},\ }\href {\doibase
  10.1103/PhysRevLett.107.060404} {\bibfield  {journal} {\bibinfo  {journal}
  {Phys. Rev. Lett.}\ }\textbf {\bibinfo {volume} {107}},\ \bibinfo {pages}
  {060404} (\bibinfo {year} {2011})}\BibitemShut {NoStop}%
\bibitem [{\citenamefont {Sinatra}\ \emph {et~al.}(2013)\citenamefont
  {Sinatra}, \citenamefont {Castin},\ and\ \citenamefont
  {Witkowska}}]{Sinatra13b}%
  \BibitemOpen
  \bibfield  {author} {\bibinfo {author} {\bibfnamefont {A.}~\bibnamefont
  {Sinatra}}, \bibinfo {author} {\bibfnamefont {Y.}~\bibnamefont {Castin}}, \
  and\ \bibinfo {author} {\bibfnamefont {E.}~\bibnamefont {Witkowska}},\ }\href
  {http://stacks.iop.org/0295-5075/102/i=4/a=40001} {\bibfield  {journal}
  {\bibinfo  {journal} {EPL (Europhysics Letters)}\ }\textbf {\bibinfo {volume}
  {102}},\ \bibinfo {pages} {40001} (\bibinfo {year} {2013})}\BibitemShut
  {NoStop}%
\bibitem [{\citenamefont {Gati}\ \emph {et~al.}(2006)\citenamefont {Gati},
  \citenamefont {Esteve}, \citenamefont {Hemmerling}, \citenamefont
  {Ottenstein}, \citenamefont {Appmeier}, \citenamefont {Weller},\ and\
  \citenamefont {Oberthaler}}]{Gati06}%
  \BibitemOpen
  \bibfield  {author} {\bibinfo {author} {\bibfnamefont {R.}~\bibnamefont
  {Gati}}, \bibinfo {author} {\bibfnamefont {J.}~\bibnamefont {Esteve}},
  \bibinfo {author} {\bibfnamefont {B.}~\bibnamefont {Hemmerling}}, \bibinfo
  {author} {\bibfnamefont {T.~B.}\ \bibnamefont {Ottenstein}}, \bibinfo
  {author} {\bibfnamefont {J.}~\bibnamefont {Appmeier}}, \bibinfo {author}
  {\bibfnamefont {A.}~\bibnamefont {Weller}}, \ and\ \bibinfo {author}
  {\bibfnamefont {M.}~\bibnamefont {Oberthaler}},\ }\href
  {http://stacks.iop.org/1367-2630/8/i=9/a=189} {\bibfield  {journal} {\bibinfo
   {journal} {New Journal of Physics}\ }\textbf {\bibinfo {volume} {8}},\
  \bibinfo {pages} {189} (\bibinfo {year} {2006})}\BibitemShut {NoStop}%
\bibitem [{\citenamefont {Hofferberth}\ \emph {et~al.}(2008)\citenamefont
  {Hofferberth}, \citenamefont {Lesanovsky}, \citenamefont {Schumm},
  \citenamefont {Imambekov}, \citenamefont {Gritsev}, \citenamefont {Demler},\
  and\ \citenamefont {Schmiedmayer}}]{Hofferberth08}%
  \BibitemOpen
  \bibfield  {author} {\bibinfo {author} {\bibfnamefont {S.}~\bibnamefont
  {Hofferberth}}, \bibinfo {author} {\bibfnamefont {I.}~\bibnamefont
  {Lesanovsky}}, \bibinfo {author} {\bibfnamefont {T.}~\bibnamefont {Schumm}},
  \bibinfo {author} {\bibfnamefont {A.}~\bibnamefont {Imambekov}}, \bibinfo
  {author} {\bibfnamefont {V.}~\bibnamefont {Gritsev}}, \bibinfo {author}
  {\bibfnamefont {E.}~\bibnamefont {Demler}}, \ and\ \bibinfo {author}
  {\bibfnamefont {J.}~\bibnamefont {Schmiedmayer}},\ }\href@noop {} {\ \textbf
  {\bibinfo {volume} {4}},\ \bibinfo {pages} {489 } (\bibinfo {year}
  {2008})}\BibitemShut {NoStop}%
\bibitem [{\citenamefont {Dalfovo}\ \emph {et~al.}(1999)\citenamefont
  {Dalfovo}, \citenamefont {Giorgini}, \citenamefont {Pitaevskii},\ and\
  \citenamefont {Stringari}}]{Dalfovo99}%
  \BibitemOpen
  \bibfield  {author} {\bibinfo {author} {\bibfnamefont {F.}~\bibnamefont
  {Dalfovo}}, \bibinfo {author} {\bibfnamefont {S.}~\bibnamefont {Giorgini}},
  \bibinfo {author} {\bibfnamefont {L.~P.}\ \bibnamefont {Pitaevskii}}, \ and\
  \bibinfo {author} {\bibfnamefont {S.}~\bibnamefont {Stringari}},\ }\href
  {\doibase 10.1103/RevModPhys.71.463} {\bibfield  {journal} {\bibinfo
  {journal} {Rev. Mod. Phys.}\ }\textbf {\bibinfo {volume} {71}},\ \bibinfo
  {pages} {463} (\bibinfo {year} {1999})}\BibitemShut {NoStop}%
\bibitem [{\citenamefont {Jin}\ \emph {et~al.}(1996)\citenamefont {Jin},
  \citenamefont {Ensher}, \citenamefont {Matthews}, \citenamefont {Wieman},\
  and\ \citenamefont {Cornell}}]{Jin96}%
  \BibitemOpen
  \bibfield  {author} {\bibinfo {author} {\bibfnamefont {D.~S.}\ \bibnamefont
  {Jin}}, \bibinfo {author} {\bibfnamefont {J.~R.}\ \bibnamefont {Ensher}},
  \bibinfo {author} {\bibfnamefont {M.~R.}\ \bibnamefont {Matthews}}, \bibinfo
  {author} {\bibfnamefont {C.~E.}\ \bibnamefont {Wieman}}, \ and\ \bibinfo
  {author} {\bibfnamefont {E.~A.}\ \bibnamefont {Cornell}},\ }\href {\doibase
  10.1103/PhysRevLett.77.420} {\bibfield  {journal} {\bibinfo  {journal} {Phys.
  Rev. Lett.}\ }\textbf {\bibinfo {volume} {77}},\ \bibinfo {pages} {420}
  (\bibinfo {year} {1996})}\BibitemShut {NoStop}%
\bibitem [{\citenamefont {Mewes}\ \emph {et~al.}(1996)\citenamefont {Mewes},
  \citenamefont {Andrews}, \citenamefont {van Druten}, \citenamefont {Kurn},
  \citenamefont {Durfee}, \citenamefont {Townsend},\ and\ \citenamefont
  {Ketterle}}]{Mewes96}%
  \BibitemOpen
  \bibfield  {author} {\bibinfo {author} {\bibfnamefont {M.-O.}\ \bibnamefont
  {Mewes}}, \bibinfo {author} {\bibfnamefont {M.~R.}\ \bibnamefont {Andrews}},
  \bibinfo {author} {\bibfnamefont {N.~J.}\ \bibnamefont {van Druten}},
  \bibinfo {author} {\bibfnamefont {D.~M.}\ \bibnamefont {Kurn}}, \bibinfo
  {author} {\bibfnamefont {D.~S.}\ \bibnamefont {Durfee}}, \bibinfo {author}
  {\bibfnamefont {C.~G.}\ \bibnamefont {Townsend}}, \ and\ \bibinfo {author}
  {\bibfnamefont {W.}~\bibnamefont {Ketterle}},\ }\href {\doibase
  10.1103/PhysRevLett.77.988} {\bibfield  {journal} {\bibinfo  {journal} {Phys.
  Rev. Lett.}\ }\textbf {\bibinfo {volume} {77}},\ \bibinfo {pages} {988}
  (\bibinfo {year} {1996})}\BibitemShut {NoStop}%
\bibitem [{\citenamefont {Jin}\ \emph {et~al.}(1997)\citenamefont {Jin},
  \citenamefont {Matthews}, \citenamefont {Ensher}, \citenamefont {Wieman},\
  and\ \citenamefont {Cornell}}]{Jin97}%
  \BibitemOpen
  \bibfield  {author} {\bibinfo {author} {\bibfnamefont {D.~S.}\ \bibnamefont
  {Jin}}, \bibinfo {author} {\bibfnamefont {M.~R.}\ \bibnamefont {Matthews}},
  \bibinfo {author} {\bibfnamefont {J.~R.}\ \bibnamefont {Ensher}}, \bibinfo
  {author} {\bibfnamefont {C.~E.}\ \bibnamefont {Wieman}}, \ and\ \bibinfo
  {author} {\bibfnamefont {E.~A.}\ \bibnamefont {Cornell}},\ }\href {\doibase
  10.1103/PhysRevLett.78.764} {\bibfield  {journal} {\bibinfo  {journal} {Phys.
  Rev. Lett.}\ }\textbf {\bibinfo {volume} {78}},\ \bibinfo {pages} {764}
  (\bibinfo {year} {1997})}\BibitemShut {NoStop}%
\bibitem [{\citenamefont {Stenger}\ \emph {et~al.}(1999)\citenamefont
  {Stenger}, \citenamefont {Inouye}, \citenamefont {Chikkatur}, \citenamefont
  {Stamper-Kurn}, \citenamefont {Pritchard},\ and\ \citenamefont
  {Ketterle}}]{Stenger99}%
  \BibitemOpen
  \bibfield  {author} {\bibinfo {author} {\bibfnamefont {J.}~\bibnamefont
  {Stenger}}, \bibinfo {author} {\bibfnamefont {S.}~\bibnamefont {Inouye}},
  \bibinfo {author} {\bibfnamefont {A.~P.}\ \bibnamefont {Chikkatur}}, \bibinfo
  {author} {\bibfnamefont {D.~M.}\ \bibnamefont {Stamper-Kurn}}, \bibinfo
  {author} {\bibfnamefont {D.~E.}\ \bibnamefont {Pritchard}}, \ and\ \bibinfo
  {author} {\bibfnamefont {W.}~\bibnamefont {Ketterle}},\ }\href {\doibase
  10.1103/PhysRevLett.82.4569} {\bibfield  {journal} {\bibinfo  {journal}
  {Phys. Rev. Lett.}\ }\textbf {\bibinfo {volume} {82}},\ \bibinfo {pages}
  {4569} (\bibinfo {year} {1999})}\BibitemShut {NoStop}%
\bibitem [{\citenamefont {Stamper-Kurn}\ \emph {et~al.}(1999)\citenamefont
  {Stamper-Kurn}, \citenamefont {Chikkatur}, \citenamefont {G\"orlitz},
  \citenamefont {Inouye}, \citenamefont {Gupta}, \citenamefont {Pritchard},\
  and\ \citenamefont {Ketterle}}]{Stamper99}%
  \BibitemOpen
  \bibfield  {author} {\bibinfo {author} {\bibfnamefont {D.~M.}\ \bibnamefont
  {Stamper-Kurn}}, \bibinfo {author} {\bibfnamefont {A.~P.}\ \bibnamefont
  {Chikkatur}}, \bibinfo {author} {\bibfnamefont {A.}~\bibnamefont
  {G\"orlitz}}, \bibinfo {author} {\bibfnamefont {S.}~\bibnamefont {Inouye}},
  \bibinfo {author} {\bibfnamefont {S.}~\bibnamefont {Gupta}}, \bibinfo
  {author} {\bibfnamefont {D.~E.}\ \bibnamefont {Pritchard}}, \ and\ \bibinfo
  {author} {\bibfnamefont {W.}~\bibnamefont {Ketterle}},\ }\href {\doibase
  10.1103/PhysRevLett.83.2876} {\bibfield  {journal} {\bibinfo  {journal}
  {Phys. Rev. Lett.}\ }\textbf {\bibinfo {volume} {83}},\ \bibinfo {pages}
  {2876} (\bibinfo {year} {1999})}\BibitemShut {NoStop}%
\bibitem [{\citenamefont {Ozeri}\ \emph {et~al.}(2005)\citenamefont {Ozeri},
  \citenamefont {Katz}, \citenamefont {Steinhauer},\ and\ \citenamefont
  {Davidson}}]{Ozeri05}%
  \BibitemOpen
  \bibfield  {author} {\bibinfo {author} {\bibfnamefont {R.}~\bibnamefont
  {Ozeri}}, \bibinfo {author} {\bibfnamefont {N.}~\bibnamefont {Katz}},
  \bibinfo {author} {\bibfnamefont {J.}~\bibnamefont {Steinhauer}}, \ and\
  \bibinfo {author} {\bibfnamefont {N.}~\bibnamefont {Davidson}},\ }\href
  {\doibase 10.1103/RevModPhys.77.187} {\bibfield  {journal} {\bibinfo
  {journal} {Rev. Mod. Phys.}\ }\textbf {\bibinfo {volume} {77}},\ \bibinfo
  {pages} {187} (\bibinfo {year} {2005})}\BibitemShut {NoStop}%
\bibitem [{\citenamefont {Diedrich}\ \emph {et~al.}(1989)\citenamefont
  {Diedrich}, \citenamefont {Bergquist}, \citenamefont {Itano},\ and\
  \citenamefont {Wineland}}]{Diedrich89}%
  \BibitemOpen
  \bibfield  {author} {\bibinfo {author} {\bibfnamefont {F.}~\bibnamefont
  {Diedrich}}, \bibinfo {author} {\bibfnamefont {J.~C.}\ \bibnamefont
  {Bergquist}}, \bibinfo {author} {\bibfnamefont {W.~M.}\ \bibnamefont
  {Itano}}, \ and\ \bibinfo {author} {\bibfnamefont {D.~J.}\ \bibnamefont
  {Wineland}},\ }\href {\doibase 10.1103/PhysRevLett.62.403} {\bibfield
  {journal} {\bibinfo  {journal} {Phys. Rev. Lett.}\ }\textbf {\bibinfo
  {volume} {62}},\ \bibinfo {pages} {403} (\bibinfo {year} {1989})}\BibitemShut
  {NoStop}%
\bibitem [{\citenamefont {Wineland}\ \emph {et~al.}(1998)\citenamefont
  {Wineland}, \citenamefont {Monroe}, \citenamefont {Itano}, \citenamefont
  {Leibfreid}, \citenamefont {King},\ and\ \citenamefont
  {Meekhof}}]{Wineland98}%
  \BibitemOpen
  \bibfield  {author} {\bibinfo {author} {\bibfnamefont {D.~J.}\ \bibnamefont
  {Wineland}}, \bibinfo {author} {\bibfnamefont {C.}~\bibnamefont {Monroe}},
  \bibinfo {author} {\bibfnamefont {W.}~\bibnamefont {Itano}}, \bibinfo
  {author} {\bibfnamefont {D.}~\bibnamefont {Leibfreid}}, \bibinfo {author}
  {\bibfnamefont {B.}~\bibnamefont {King}}, \ and\ \bibinfo {author}
  {\bibfnamefont {D.~M.}\ \bibnamefont {Meekhof}},\ }\href
  {http://tf.nist.gov/general/pdf/1275.pdf} {\bibfield  {journal} {\bibinfo
  {journal} {J. Res. Natl. Inst. Stand. Technol.}\ }\textbf {\bibinfo {volume}
  {103}},\ \bibinfo {pages} {259} (\bibinfo {year} {1998})}\BibitemShut
  {NoStop}%
\bibitem [{\citenamefont {Boozer}\ \emph {et~al.}(2006)\citenamefont {Boozer},
  \citenamefont {Boca}, \citenamefont {Miller}, \citenamefont {Northup},\ and\
  \citenamefont {Kimble}}]{Boozer06}%
  \BibitemOpen
  \bibfield  {author} {\bibinfo {author} {\bibfnamefont {A.~D.}\ \bibnamefont
  {Boozer}}, \bibinfo {author} {\bibfnamefont {A.}~\bibnamefont {Boca}},
  \bibinfo {author} {\bibfnamefont {R.}~\bibnamefont {Miller}}, \bibinfo
  {author} {\bibfnamefont {T.~E.}\ \bibnamefont {Northup}}, \ and\ \bibinfo
  {author} {\bibfnamefont {H.~J.}\ \bibnamefont {Kimble}},\ }\href {\doibase
  10.1103/PhysRevLett.97.083602} {\bibfield  {journal} {\bibinfo  {journal}
  {Phys. Rev. Lett.}\ }\textbf {\bibinfo {volume} {97}},\ \bibinfo {pages}
  {083602} (\bibinfo {year} {2006})}\BibitemShut {NoStop}%
\bibitem [{\citenamefont {Kaufman}\ \emph {et~al.}(2012)\citenamefont
  {Kaufman}, \citenamefont {Lester},\ and\ \citenamefont {Regal}}]{Kaufman12}%
  \BibitemOpen
  \bibfield  {author} {\bibinfo {author} {\bibfnamefont {A.~M.}\ \bibnamefont
  {Kaufman}}, \bibinfo {author} {\bibfnamefont {B.~J.}\ \bibnamefont {Lester}},
  \ and\ \bibinfo {author} {\bibfnamefont {C.~A.}\ \bibnamefont {Regal}},\
  }\href {\doibase 10.1103/PhysRevX.2.041014} {\bibfield  {journal} {\bibinfo
  {journal} {Phys. Rev. X}\ }\textbf {\bibinfo {volume} {2}},\ \bibinfo {pages}
  {041014} (\bibinfo {year} {2012})}\BibitemShut {NoStop}%
\bibitem [{\citenamefont {F\"orster}\ \emph {et~al.}(2009)\citenamefont
  {F\"orster}, \citenamefont {Karski}, \citenamefont {Choi}, \citenamefont
  {Steffen}, \citenamefont {Alt}, \citenamefont {Meschede}, \citenamefont
  {Widera}, \citenamefont {Montano}, \citenamefont {Lee}, \citenamefont
  {Rakreungdet},\ and\ \citenamefont {Jessen}}]{Foerster09}%
  \BibitemOpen
  \bibfield  {author} {\bibinfo {author} {\bibfnamefont {L.}~\bibnamefont
  {F\"orster}}, \bibinfo {author} {\bibfnamefont {M.}~\bibnamefont {Karski}},
  \bibinfo {author} {\bibfnamefont {J.-M.}\ \bibnamefont {Choi}}, \bibinfo
  {author} {\bibfnamefont {A.}~\bibnamefont {Steffen}}, \bibinfo {author}
  {\bibfnamefont {W.}~\bibnamefont {Alt}}, \bibinfo {author} {\bibfnamefont
  {D.}~\bibnamefont {Meschede}}, \bibinfo {author} {\bibfnamefont
  {A.}~\bibnamefont {Widera}}, \bibinfo {author} {\bibfnamefont
  {E.}~\bibnamefont {Montano}}, \bibinfo {author} {\bibfnamefont {J.~H.}\
  \bibnamefont {Lee}}, \bibinfo {author} {\bibfnamefont {W.}~\bibnamefont
  {Rakreungdet}}, \ and\ \bibinfo {author} {\bibfnamefont {P.~S.}\ \bibnamefont
  {Jessen}},\ }\href {\doibase 10.1103/PhysRevLett.103.233001} {\bibfield
  {journal} {\bibinfo  {journal} {Phys. Rev. Lett.}\ }\textbf {\bibinfo
  {volume} {103}},\ \bibinfo {pages} {233001} (\bibinfo {year}
  {2009})}\BibitemShut {NoStop}%
\bibitem [{\citenamefont {B\"ohi}\ \emph {et~al.}(2009)\citenamefont {B\"ohi},
  \citenamefont {Riedel}, \citenamefont {Hoffrogge}, \citenamefont {Reichel},
  \citenamefont {H\"ansch},\ and\ \citenamefont {Treutlein}}]{Boehi09}%
  \BibitemOpen
  \bibfield  {author} {\bibinfo {author} {\bibfnamefont {P.}~\bibnamefont
  {B\"ohi}}, \bibinfo {author} {\bibfnamefont {M.~F.}\ \bibnamefont {Riedel}},
  \bibinfo {author} {\bibfnamefont {J.}~\bibnamefont {Hoffrogge}}, \bibinfo
  {author} {\bibfnamefont {J.}~\bibnamefont {Reichel}}, \bibinfo {author}
  {\bibfnamefont {T.~W.}\ \bibnamefont {H\"ansch}}, \ and\ \bibinfo {author}
  {\bibfnamefont {P.}~\bibnamefont {Treutlein}},\ }\href
  {http://www.nature.com/nphys/journal/v5/n8/full/nphys1329.html} {\bibfield
  {journal} {\bibinfo  {journal} {Nature Physics}\ }\textbf {\bibinfo {volume}
  {5}},\ \bibinfo {pages} {592} (\bibinfo {year} {2009})}\BibitemShut {NoStop}%
\bibitem [{\citenamefont {Treutlein}\ \emph {et~al.}(2004)\citenamefont
  {Treutlein}, \citenamefont {Hommelhoff}, \citenamefont {Steinmetz},
  \citenamefont {H\"ansch},\ and\ \citenamefont {Reichel}}]{Treutlein04}%
  \BibitemOpen
  \bibfield  {author} {\bibinfo {author} {\bibfnamefont {P.}~\bibnamefont
  {Treutlein}}, \bibinfo {author} {\bibfnamefont {P.}~\bibnamefont
  {Hommelhoff}}, \bibinfo {author} {\bibfnamefont {T.}~\bibnamefont
  {Steinmetz}}, \bibinfo {author} {\bibfnamefont {T.~W.}\ \bibnamefont
  {H\"ansch}}, \ and\ \bibinfo {author} {\bibfnamefont {J.}~\bibnamefont
  {Reichel}},\ }\href {\doibase 10.1103/PhysRevLett.92.203005} {\bibfield
  {journal} {\bibinfo  {journal} {Phys. Rev. Lett.}\ }\textbf {\bibinfo
  {volume} {92}},\ \bibinfo {pages} {203005} (\bibinfo {year}
  {2004})}\BibitemShut {NoStop}%
\bibitem [{\citenamefont {Harber}\ \emph {et~al.}(2002)\citenamefont {Harber},
  \citenamefont {Lewandowski}, \citenamefont {McGuirk},\ and\ \citenamefont
  {Cornell}}]{Harber02}%
  \BibitemOpen
  \bibfield  {author} {\bibinfo {author} {\bibfnamefont {D.~M.}\ \bibnamefont
  {Harber}}, \bibinfo {author} {\bibfnamefont {H.~J.}\ \bibnamefont
  {Lewandowski}}, \bibinfo {author} {\bibfnamefont {J.~M.}\ \bibnamefont
  {McGuirk}}, \ and\ \bibinfo {author} {\bibfnamefont {E.~A.}\ \bibnamefont
  {Cornell}},\ }\href {\doibase 10.1103/PhysRevA.66.053616} {\bibfield
  {journal} {\bibinfo  {journal} {Phys. Rev. A}\ }\textbf {\bibinfo {volume}
  {66}},\ \bibinfo {pages} {053616} (\bibinfo {year} {2002})}\BibitemShut
  {NoStop}%
\bibitem [{\citenamefont {Ketterle}\ \emph {et~al.}(1999)\citenamefont
  {Ketterle}, \citenamefont {Durfee},\ and\ \citenamefont {Kurn}}]{Ketterle99}%
  \BibitemOpen
  \bibfield  {author} {\bibinfo {author} {\bibfnamefont {W.}~\bibnamefont
  {Ketterle}}, \bibinfo {author} {\bibfnamefont {D.~S.}\ \bibnamefont
  {Durfee}}, \ and\ \bibinfo {author} {\bibfnamefont {S.~D.~M.}\ \bibnamefont
  {Kurn}},\ }in\ \href@noop {} {\emph {\bibinfo {booktitle} {Bose-Einstein
  Condensation in Atomic Gases (Proceedings of the International School of
  Physics ``Enrico Fermi,'' Course CXL)}}},\ \bibinfo {editor} {edited by\
  \bibinfo {editor} {\bibfnamefont {M.}~\bibnamefont {Inguscio}}, \bibinfo
  {editor} {\bibfnamefont {S.}~\bibnamefont {Stringari}}, \ and\ \bibinfo
  {editor} {\bibfnamefont {C.~E.}\ \bibnamefont {Wieman}}}\ (\bibinfo
  {publisher} {IOS Press},\ \bibinfo {year} {1999})\BibitemShut {NoStop}%
\bibitem [{\citenamefont {Gehm}\ \emph {et~al.}(1998)\citenamefont {Gehm},
  \citenamefont {O'Hara}, \citenamefont {Savard},\ and\ \citenamefont
  {Thomas}}]{Gehm98}%
  \BibitemOpen
  \bibfield  {author} {\bibinfo {author} {\bibfnamefont {M.~E.}\ \bibnamefont
  {Gehm}}, \bibinfo {author} {\bibfnamefont {K.~M.}\ \bibnamefont {O'Hara}},
  \bibinfo {author} {\bibfnamefont {T.~A.}\ \bibnamefont {Savard}}, \ and\
  \bibinfo {author} {\bibfnamefont {J.~E.}\ \bibnamefont {Thomas}},\ }\href
  {\doibase 10.1103/PhysRevA.58.3914} {\bibfield  {journal} {\bibinfo
  {journal} {Phys. Rev. A}\ }\textbf {\bibinfo {volume} {58}},\ \bibinfo
  {pages} {3914} (\bibinfo {year} {1998})}\BibitemShut {NoStop}%
\bibitem [{Note1()}]{Note1}%
  \BibitemOpen
  \bibinfo {note} {Because $\Omega _{\protect \mathrm {Rabi}}\ll \omega
  _{x,y,z}$, the excited-state energies and mode functions change adiabatically
  as the mean-field potential changes and do not acquire a spatial
  dependence.}\BibitemShut {Stop}%
\bibitem [{Note2()}]{Note2}%
  \BibitemOpen
  \bibinfo {note} {Time of flight images confirm that the temperature of such a
  heated cloud is above the BEC transition temperature.}\BibitemShut {Stop}%
\bibitem [{\citenamefont {Egorov}\ \emph {et~al.}(2013)\citenamefont {Egorov},
  \citenamefont {Opanchuk}, \citenamefont {Drummond}, \citenamefont {Hall},
  \citenamefont {Hannaford},\ and\ \citenamefont {Sidorov}}]{Egorov13}%
  \BibitemOpen
  \bibfield  {author} {\bibinfo {author} {\bibfnamefont {M.}~\bibnamefont
  {Egorov}}, \bibinfo {author} {\bibfnamefont {B.}~\bibnamefont {Opanchuk}},
  \bibinfo {author} {\bibfnamefont {P.}~\bibnamefont {Drummond}}, \bibinfo
  {author} {\bibfnamefont {B.~V.}\ \bibnamefont {Hall}}, \bibinfo {author}
  {\bibfnamefont {P.}~\bibnamefont {Hannaford}}, \ and\ \bibinfo {author}
  {\bibfnamefont {A.~I.}\ \bibnamefont {Sidorov}},\ }\href {\doibase
  10.1103/PhysRevA.87.053614} {\bibfield  {journal} {\bibinfo  {journal} {Phys.
  Rev. A}\ }\textbf {\bibinfo {volume} {87}},\ \bibinfo {pages} {053614}
  (\bibinfo {year} {2013})}\BibitemShut {NoStop}%
\bibitem [{\citenamefont {Cirac}\ \emph {et~al.}(1998)\citenamefont {Cirac},
  \citenamefont {Lewenstein}, \citenamefont {M\o{}lmer},\ and\ \citenamefont
  {Zoller}}]{Cirac98}%
  \BibitemOpen
  \bibfield  {author} {\bibinfo {author} {\bibfnamefont {J.~I.}\ \bibnamefont
  {Cirac}}, \bibinfo {author} {\bibfnamefont {M.}~\bibnamefont {Lewenstein}},
  \bibinfo {author} {\bibfnamefont {K.}~\bibnamefont {M\o{}lmer}}, \ and\
  \bibinfo {author} {\bibfnamefont {P.}~\bibnamefont {Zoller}},\ }\href
  {\doibase 10.1103/PhysRevA.57.1208} {\bibfield  {journal} {\bibinfo
  {journal} {Phys. Rev. A}\ }\textbf {\bibinfo {volume} {57}},\ \bibinfo
  {pages} {1208} (\bibinfo {year} {1998})}\BibitemShut {NoStop}%
\bibitem [{\citenamefont {\"{O}hberg}\ and\ \citenamefont
  {Stenholm}(1997)}]{Oehberg97}%
  \BibitemOpen
  \bibfield  {author} {\bibinfo {author} {\bibfnamefont {P.}~\bibnamefont
  {\"{O}hberg}}\ and\ \bibinfo {author} {\bibfnamefont {S.}~\bibnamefont
  {Stenholm}},\ }\href {http://stacks.iop.org/0953-4075/30/i=12/a=005}
  {\bibfield  {journal} {\bibinfo  {journal} {Journal of Physics B: Atomic,
  Molecular and Optical Physics}\ }\textbf {\bibinfo {volume} {30}},\ \bibinfo
  {pages} {2749} (\bibinfo {year} {1997})}\BibitemShut {NoStop}%
\bibitem [{Note3()}]{Note3}%
  \BibitemOpen
  \bibinfo {note} {There is another line corresponding to the transition from
  the first excited vibrational state in $\vert 1 \rangle $ ($\alpha =1$) to
  the lowest vibrational state in $\vert 2 \rangle $($\beta =0$) at an offset
  of $-240~\protect \text {Hz}$ from the carrier. This transition is weaker
  because it has a lower wavefunction overlap. We focus our analysis on the
  strongest transition involving the thermal cloud.}\BibitemShut {Stop}%
\bibitem [{Note4()}]{Note4}%
  \BibitemOpen
  \bibinfo {note} {We point out that in the specific case of a BEC in a
  double-well potential, a primary noise thermometer is presented in
  Ref~\protect \rev@citealp {Gati06}.}\BibitemShut {Stop}%
\end{thebibliography}%

\end{document}